\documentclass[aps,pra,twocolumn]{revtex4}

\usepackage{amsmath,amsthm,amssymb,url,graphicx}
\usepackage{dcolumn}
\usepackage[latin1]{inputenc}
\usepackage[francais,english]{babel}

\newcommand{\R}{{\mathbb{R}}}

\newcommand{\di}{{\rm d}}
\newcommand{\ee}{{\rm e}}

\newcommand{\eps}{\epsilon}

\newcommand{\fv}{{\bf f}}

\newcommand{\rv}{{\bf r}}
\newcommand{\sv}{{\bf s}}

\newcommand{\bmath}{\begin{eqnarray}}
\newcommand{\emath}{\end{eqnarray}}

\begin{document}
	
\title{Challenging the Lieb-Oxford bound in a systematic way}	
\author{Michael Seidl,$^{1,2}$ Stefan Vuckovic,$^2$ and Paola Gori-Giorgi$^2$}
\affiliation{$^1$Institute of Theoretical Physics, University of Regensburg, D-93040 Regensburg, Germany \\
$^2$Department of Theoretical Chemistry and Amsterdam Center for Multiscale Modeling, FEW, Vrije Universiteit, De Boelelaan 1083, 1081HV Amsterdam, The Netherlands}

\begin{abstract}
The Lieb-Oxford bound, a nontrivial inequality for the indirect part of the many-body Coulomb repulsion in an electronic system, plays an important role in the construction of approximations in density functional theory.
Using the wavefunction for strictly-correlated electrons of a given density, we turn the search over wavefunctions appearing in the original bound into a more manageable search over electron densities. This allows us to challenge the bound in a systematic way. We find that a maximizing density for the bound, if it exists, must have compact support. We also find that, at least for particle numbers $N\le 60$, a uniform density profile is not the most challenging for the bound. With our construction we improve the bound for $N=2$ electrons that was originally found by Lieb and Oxford, we give a new lower bound to the constant appearing in the Lieb-Oxford inequality valid for any $N$, and we provide an improved upper bound for the low-density uniform electron gas indirect energy.
\end{abstract}

\maketitle

\section{Introduction}
Lieb and Oxford (LO) \cite{Lie-PLA-79,LieOxf-IJQC-81} proved a nontrivial inequality for the indirect part of the electron-electron interaction energy (total expectation of the interaction minus the Hartree term) with respect to the LDA exchange functional. This inequality has been recently extended to include the gradient of the density \cite{LewLie-PRA-15}. 
The LO bound has played and continues to play a very important role in the construction of approximate exchange-correlation (xc) density functionals \cite{LevPer-PRB-93,OdaCap-JCP-07,OdaCap-IJQC-08,OdaCap-PRA-09,OdaCapTri-JCTC-09,HauOdaScuPerCap-JCP-12,PerRuzTaoStaScuCso-JCP-05,PerRuzSunBur-JCP-14,SunRuzPer-PRL-15,ConTerDelFab-PRB-15}.
While traditionally only the more general LO bound, valid for any number of particles $N$ (and corresponding to $N\to
\infty$) has been taken into account in the construction of xc approximations, it has been shown very recently that the bound for $N=1$ and $N=2$ is important in the context of metaGGA functionals \cite{PerRuzSunBur-JCP-14,SunRuzPer-PRL-15}, and can be imposed as an additional exact condition.

The bound for $N=1$ was first given in Ref.~\onlinecite{GadBarHan-JCP-80}, and proved rigorously in Ref.~\onlinecite{LieOxf-IJQC-81}. For $N=2$, Lieb and Oxford \cite{LieOxf-IJQC-81} could only provide a non optimal estimate of the constant appearing in the bound. In this work we develop a strategy to systematically challenge the original LO bound for a given number of electrons $N$. We use optimal trial wave functions for a given density, and we then vary the density in order to challenge the bound as much as possible. After showing that a density that maximally challenges the bound, if it exists, must have compact support, we follow the functional derivative of the bound to challenge it as much as possible without violating $N$-representability also for densities whose support is the whole space. As a first application of this procedure, we improve the lower bound for $N=2$ given by Lieb and Oxford, see Eq.~\eqref{eq_lambda2final} below. Our construction also provides an improved lower bound for the constant appearing in the Lieb-Oxford inequality valid for any $N$, see Eqs.~\eqref{eq_newlower}-\eqref{eq_newC}, and an improved upper bound for the indirect energy if the low-density uniform electron gas, see Eq.~\eqref{eq_jelllowdens}.

\subsection{Notation}

In electronic density functional theory (DFT) one is interested in finding the ground-state energy and density of $N$-electron systems with Hamiltonian
\begin{equation}
\hat{H}=\hat{T}+\hat{V}_{\rm ee}+\hat{V}_{\rm ext},\qquad\hat{V}_{\rm ext}=\sum_{i=1}^Nv(\rv_i).
\label{eq:H}
\end{equation}
$\hat{T}$ and $\hat{V}_{\rm ee}$ are, respectively, the universal operators of the kinetic energy (in Hartree atomic units used throughout the paper),
\begin{equation}
\hat{T}=-\frac{1}{2}\sum_{i=1}^N\frac{\partial^2}{\partial\rv_i^2},
\end{equation}
and of the interaction (Coulomb repulsion) energy between the $N$ electrons,
\begin{equation}
\hat{V}_{\rm ee}=\frac12\sum_{i,j=1}^N\frac{1-\delta_{ij}}{|\rv_i-\rv_j|}.
\label{VeeDEF}
\end{equation}
The function $v(\rv)$, in contrast, is a non-universal but arbitrary attractive external potential
required to bind the repulsive electrons. Most of the formalism will be carried out for general spatial dimension $D=2$ and 3, $\rv\in\R^D$, focussing later on $D=3$ only.

In the following, $\Psi$ denotes a correctly normalized and antisymmetrized, but otherwise arbitrary
$N$-electron wave function (thus not necessarily eigenstate of \eqref{eq:H}),
\begin{equation}
\Psi=\Psi(\rv_1\sigma_1,...,\rv_N\sigma_N),
\end{equation}
where $\sigma_n$ are spin variables.
By $\rho_\Psi$, we denote the particle density associated with $\Psi$,
\begin{equation}
\rho_\Psi(\rv)=N\sum_{\sigma_i}\int d^Dr_2...d^Dr_N
\big|\Psi(\rv\sigma_1,\rv_2\sigma_2,...,\rv_N\sigma_N)\big|^2.
\end{equation}

\subsection{Indirect Coulomb energy}

The electronic interaction energy in the quantum state $\Psi$, defined as the expectation
\begin{equation}
\langle\Psi|\hat{V}_{\rm ee}|\Psi\rangle>0,
\end{equation}
excludes the infinite self energies of the point electrons, see the factor $1-\delta_{ij}$
in Eq.~\eqref{VeeDEF}. If the electrons were a classical continuous distribution of
negative charge with density $\rho_\Psi(\rv)$, their interaction energy would be $U[\rho_\Psi]$,
with the Hartree functional
\begin{equation}
U[\rho]=\frac12\int d^Dr\int d^Dr'\,\frac{\rho(\rv)\rho(\rv')}{|\rv-\rv'|}>0.
\end{equation}
The most severe error introduced by this classical continuum approximation is a spurious finite self-interaction
energy included for each electron. This is particularly evident in the case $N=1$, since for any normalized
one-electron wave function $\Psi$, we have $\langle\Psi|\hat{V}_{\rm ee}|\Psi\rangle=0$, while $U[\rho_\Psi]>0$.
The indirect interaction energy $W[\Psi]$ is defined as
\begin{equation}
W[\Psi]\;\equiv\;\langle\Psi|\hat{V}_{\rm ee}|\Psi\rangle-U[\rho_\Psi].
\end{equation}
For wavefunctions that are ground states of an $N$-electron hamiltonian \eqref{eq:H} (or good trial wavefunction for it) $W[\Psi]$ is normally negative. However, for a given density $\rho$, it is possible to construct  wavefunctions $\Psi$ for which $W[\Psi]$ is positive or even infinity \cite{LieOxf-IJQC-81,SeiGor-PRA-10}.
We emphasize that $U[\rho]$ is a density functional, while $W[\Psi]$
is a functional in terms of the wave function $\Psi$.

\subsection{Lieb-Oxford bound}
The quantity $W[\Psi]$ is limited by the Lieb-Oxford (LO) bound,
\begin{equation}
-C_D\int d^Dr\,\rho_\Psi(\rv)^{1+1/D}\;\le\;W[\Psi].
\label{LOorg}
\end{equation}
$C_D>0$ is the unknown minimum possible number that makes this inequality true for all wave functions $\Psi$ in $D=2$ or 3 dimensions. So far, it is rigorously known that  $C_3\le1.6358$ \cite{ChaHan-PRA-99} and  $C_2\le 481.28$ \cite{LieSolYng-PRB-95}, and it has been argued \cite{RasPitCapPro-PRL-09}, on physical arguments, that the two bounds can be tightened to $C_3\le1.44$ and $C_2\le1.96$. The assumption behind these latter conjectured values is that the tightest possible bound is provided by the indirect energy of the uniform electron gas in the low-density limit, which, in turn, is commonly identified with the Wigner crystal total energy. This latter assumption has recently been proven wrong for the 3D case by Lewin and Lieb \cite{LewLie-PRA-15}. The study presented in this paper will also raise doubts on the first assumption that a uniform density is really the most challenging case for the LO bound, after a suitable optimal wave function for each given density has been defined (see Sec.~\ref{sec_unif}).

In terms of the local-density approximation (LDA)
\begin{equation}
-A_D\int d^Dr\,\rho(\rv)^{1+1/D}=E_{\rm x}^{\rm LDA}[\rho]
\end{equation}
to the $D$-dimensional exchange energy, with the exact constants $A_3=\frac34(\frac3{\pi})^{1/3}\approx 0.739$,
$A_2=\frac43(\frac2{\pi})^{1/2}\approx 0.798$, Eq.~\eqref{LOorg} reads
\begin{equation}
\lambda[\Psi]\;\le\;\bar{\lambda}_D,
\label{LOx}
\end{equation}
where we have defined
\begin{equation}
\lambda[\Psi]\equiv\frac{W[\Psi]}{E_{\rm x}^{\rm LDA}[\rho_\Psi]},\qquad
\bar{\lambda}_D\equiv\frac{C_D}{A_D}.
\end{equation}
Considering all antisymmetric wave functions $\Psi$ in $D$ dimensions, we may write
\begin{equation}
\bar{\lambda}_D\;=\;\sup_{\Psi:D}\lambda[\Psi].
\label{lambdaBar}
\end{equation}
The above rigorous upper bounds for $C_D$ correspond to 
\begin{equation}
\bar{\lambda}_2\le 603,\qquad\bar{\lambda}_3\le 2.215.
\end{equation}
Considering wave functions $\Psi\to N$ with a given particle number $N$, we define
\begin{equation}
\bar{\lambda}_D(N)=\sup_{(\Psi:D)\to N}\lambda[\Psi].
\end{equation}
Lieb and Oxford \cite{LieOxf-IJQC-81} have proven that $\bar{\lambda}_3(N)$ is monotonically increasing with its integer variable $N$,
\begin{equation}
\bar{\lambda}_3(N)<\bar{\lambda}_3(N+1),\qquad\lim_{N\to\infty}\bar{\lambda}_3(N)=\bar{\lambda}_3.
\label{eq_incN}
\end{equation}
They have also proven that  $\bar{\lambda}_3(1)=1.4786$ (which was given originally by Gadre {\em et al.} \cite{GadBarHan-JCP-80}) and they have found a lower bound for $\bar{\lambda}_3(2)$,
\begin{equation}
\bar{\lambda}_3(2)>1.67.
\label{lambdaEst0}
\end{equation}
These bounds in $D=3$ for $N=1$ and $N=2$ have been recently used to improve a certain class of exchange-correlation functionals \cite{PerRuzSunBur-JCP-14,SunRuzPer-PRL-15}.

In this paper we develop a general strategy to find improved {\em lower} bounds for $\bar{\lambda}_D(N)$ by {\em challenging} the Lieb-Oxford bound, {\em i.e}, by evaluating $\lambda[\Psi]$ with particularly efficient trial wave functions $\Psi$. Notice that this is different from what is usually called {\em tightening} the bound, which means finding improved {\em upper} bounds to $\bar{\lambda}_D(N)$. 

A new lower bound for  $\bar{\lambda}_D(N)$ (or, generally, for $\bar{\lambda}_D$) is rigorously obtained each time we find a wavefunction that gives the highest value ever observed for $\lambda[\Psi]$ (for a given $N$, or in general). 
Until very recently, it was believed that a lower bound for $\bar{\lambda}_3$ is given by $\bar{\lambda}_3\ge 1.444/A_3=1.955$, corresponding to the total energy of the bcc Wigner crystal in the classical jellium model. However, in the jellium model, one can only identify the total energy with the indirect energy if the electronic density is uniform, exactly equal to the one of the positive background. Only in this case the electronic Hartree term will be exactly canceled by the electron-background and the background-background contributions to the total energy. Lewin and Lieb \cite{LewLie-PRA-15} have shown that in the 3D case trying to make this cancellation happen by taking a superposition of all the possible Wigner lattices to have a uniform electronic density, introduces a shift that does not disappear in the thermodynamic limit. Thus, the value 1.955 does not correspond to the indirect energy of any wave function and is not a valid lower bound for $\bar{\lambda}_3$. In Sec.~\ref{sec_unif} we report a new lower bound for general $N$, by considering an optimal trial wave function for $N=60$, and we also report an improved upper bound to the indirect energy of the low-density uniform gas.

\section{The density functional $\Lambda[\rho]$}
\label{DFLambda}
Considering only those wave functions $\Psi\to\rho$ (in $D$ dimensions) that are associated with a given particle
density $\rho=\rho(\rv)$, we define the density functional
\begin{equation}
\Lambda[\rho]\;\equiv\;\max_{\Psi\to\rho}\lambda[\Psi].
\label{Lambda}
\end{equation}
Writing $N_\Psi=\int d^Dr\,\rho_\Psi(\rv)$ for the electron number in the state $\Psi$, we then have
\begin{equation}
\lambda[\Psi]\;\le\;\Lambda[\rho_\Psi]\;<\;\bar{\lambda}_D(N_\Psi)\;<\;\bar{\lambda}_D.
\end{equation}

\subsection{SCE interaction energy}

More explicitly, 
\begin{equation}
\Lambda[\rho]\;\equiv\;
\max_{\Psi\to\rho}\frac{\langle\Psi|\hat{V}_{\rm ee}|\Psi\rangle-U[\rho]}{E_{\rm x}^{\rm LDA}[\rho]}\;=\;
\frac{V_{\rm ee}^{\rm SCE}[\rho]-U[\rho]}{E_{\rm x}^{\rm LDA}[\rho]},
\label{LambdaExplic}
\end{equation}
with the SCE interaction energy of Appendix \ref{appSCE},
\begin{equation}
V_{\rm ee}^{\rm SCE}[\rho]\;=\;\min_{\Psi\to\rho}\langle\Psi|\hat{V}_{\rm ee}|\Psi\rangle.
\label{VeeSCE}
\end{equation}
The acronym SCE \cite{Sei-PRA-99,SeiPerLev-PRA-99,SeiGorSav-PRA-07} stands for ``strictly-correlated electrons'' and defines a
state $|\Psi_{\rm SCE}[\rho]|^2$, which is a very accurate trial wave function (actually a distribuition) for the maximizing one in Eq.~\eqref{LambdaExplic}, being exact in 1D \cite{ColDepDiM-CJM-14} for any $N$, and in any dimension for $N=2$ \cite{ButDepGor-PRA-12}. The SCE state is detailed in  Appendix \ref{appSCE}. In other words: {\em Out of all antisymmetric wave functions $\Psi$ that are
associated with a given density $\rho$, the one that provides (or is very close to) the strongest challenge to the Lieb-Oxford bound is
the SCE state $|\Psi_{\rm SCE}[\rho]|^2$.} Consequently, since $V_{\rm ee}^{\rm SCE}[\rho]$ can be evaluated rigorously for a wide
class of densities, Eq.~\eqref{VeeSCEfs} in Appendix \ref{appSCE}, we no longer need to consider different trial
wave functions $\Psi$, but only different trial densities $\rho$ instead,
\begin{equation}
\bar{\lambda}_D\;=\;\sup_{\rho:D}\Lambda[\rho],\qquad\bar{\lambda}_D(N)\;=\;\sup_{(\rho:D)\to N}\Lambda[\rho].
\end{equation}

As a preliminary step, we have used simple analytical trial spherical densities to evaluate $\Lambda[\rho]$ for $N=2$, reporting the results in Table~\ref{tab:trialrho}. We see that 
the lower bound \eqref{lambdaEst0} is readily improved to
\begin{equation}
\bar{\lambda}_3(2)>1.700 97.
\label{lambdaEst1}
\end{equation}
There is no need for considering scaled densities $\rho_\xi(\rv)\equiv\xi^D\rho(\xi\rv)$, with various values of
$\xi>0$, since $\Lambda[\rho_\xi]=\Lambda[\rho]$, see Eq.~\eqref{VeeSCEscal} in Appendix \ref{appSCE}. It is interesting to notice that, once the most challenging wave function for each given $\rho(\rv)$ is used, the densities that give the highest values of $\Lambda[\rho]$ are quite surprising. For example, a density proportional to $e^{-50\,(r-1)^2}$, consisting of a thin spherical shell, is similar to the one of the strongly-correlated limit of the Hooke's atom series. Yet, it gives a value of $\Lambda[\rho]$ which is much lower than the one obtained from the exponential density. Indeed, the strong-correlation limit of the Hooke's series is known to give $\lambda[\rho]=1.489$ \cite{OdaCapTri-JCTC-09}, again much less than what we obtain for exponential-like densities. The point is that previous works which analyzed numerically the LO bound \cite{OdaCap-JCP-07,OdaCap-IJQC-08,OdaCap-PRA-09,OdaCapTri-JCTC-09} focussed on physical hamiltonians of the kind \eqref{eq:H}, choosing $v(\rv)$ that could be particularly challenging for the bound. In that context, exponential-like densities would correspond to the large nuclear-charge limit of the He isoelectronic series, which is a weakly correlated system. With our construction, instead, we use the most challenging wave function for any given density, finding the unexpected trends of Table~\ref{tab:trialrho}. We also see that the density of a uniform sphere (``droplet'') is not particularly challenging for the bound, a feature that will be further analyzed in Sec.~\ref{sec_unif} for larger $N$.

\begin{table}
\begin{tabular}{|c|l||c|l|}\hline
$\rho(r)\propto$ & $\Lambda[\rho]$ & $\rho(r)\propto$ & $\Lambda[\rho]$ \\ \hline\hline
$e^{-10\,(r-1)^2}$ & 1.499 & $e^{-50\,(r-1)^2}$ & 1.262 \\ \hline
$(1+r)^{-4}$ & 1.562 & $e^{-r^2}$ & 1.689\\ \hline
$(1+r)^{-5}$ & 1.637 & $e^{-r}$ & 1.699~05\\ \hline
$(1+r)^{-6}$ & 1.662 & $r\,e^{-r}$ & 1.698~66\\ \hline
$(1+r)^{-7}$ & 1.674 & $r^{1/2}e^{-r}$ & 1.700~97\\ \hline
$(1+r)^{-{10}}$ & 1.687 & $r^{1/3}e^{-r}$ & 1.700~95\\ \hline
$1-r,\;\; r\le 1$ & 1.638 & $r^{-3},\;\; r\in[R_1,R_2]$ & 1.145\\ \hline
droplet & 1.498 & $\cos(r),\;\; r\le \frac{\pi}{2}$ & 1.627\\ \hline
\end{tabular}
\caption{Values $\Lambda[\rho]$ for some simple spherical two-electron trial densities $\rho(r)$ in three dimensions
($N=2$, $D=3$), obtained numerically from Eqs.~\eqref{Netwo}-\eqref{VeeSCEtwo} of Appendix \ref{appSCE}. In the last two rows we consider densities with compact support: ``droplet'' corresponds to the case of a sphere of uniform density \cite{RasSeiGor-PRB-11}, and the density proportional to $r^{-3}$ \cite{PerRuzSunBur-JCP-14} has been evaluated for $R_1=10^3$ and $R_2=10^5$.
[Atomic units are used, where $r$ is a dimensionless radial coordinate.]}
\label{tab:trialrho}
\end{table}

\subsection{Absence of a maximizing density without compact support}
\label{NonExist}
We now demonstrate that a function $\rho(\rv)$ that maximizes the functional $\Lambda[\rho]$ for a finite $N$ cannot be
a physical density, unless it has compact support. The argument is essentialy the same used by Lieb and Oxford \cite{LieOxf-IJQC-81} for $N=1$ and $N=2$. In terms of the SCE external potential of Appendix \ref{appSCE},
\begin{equation}
v_{\rm SCE}[\rho](\rv)\;\equiv\;\frac{\delta V_{\rm ee}^{\rm SCE}[\rho]}{\delta\rho(\rv)},
\end{equation}
and the Hartree potential
\begin{equation}
v_{\rm H}[\rho](\rv)\;\equiv\;\frac{\delta U[\rho]}{\delta\rho(\rv)}\;=\;\int\di^Dr'\,\frac{\rho(\rv')}{|\rv-\rv'|},
\end{equation}
we consider the Euler equation for maximizing $\Lambda[\rho]$. By writing $\rho(\rv)=p(\rv)^2$ to ensure $\rho(\rv)\ge 0$, and by varying $p(\rv)$ we obtain
\begin{eqnarray}
\left\{\frac{v_{\rm SCE}[\rho](\rv)-v_{\rm H}[\rho](\rv)}{E^{\rm LDA}_{\rm x}[\rho]}
 -\frac{V_{\rm ee}^{\rm SCE}[\rho]-U[\rho]}{E^{\rm LDA}_{\rm x}[\rho]^2}\,v^{\rm LDA}_{\rm x}[\rho](\rv)\right\} p(\rv) \nonumber \\
 =\mu \,p(\rv).\qquad
\label{EulerLam_p}
\end{eqnarray}
If $p(\rv)\neq 0$ everywhere, we obtain the Euler equation
\begin{eqnarray}
\frac{\delta\Lambda[\rho]}{\delta\rho(\rv)}& \equiv &
\frac{v_{\rm SCE}[\rho](\rv)-v_{\rm H}[\rho](\rv)}{E^{\rm LDA}_{\rm x}[\rho]} \nonumber \\
& - & 
\frac{V_{\rm ee}^{\rm SCE}[\rho]-U[\rho]}{E^{\rm LDA}_{\rm x}[\rho]^2}\,v^{\rm LDA}_{\rm x}[\rho](\rv)=\mu,
\label{EulerLam}
\end{eqnarray}
where $\mu$ is the Lagrange multiplier ensuring fixed particle number $N=\int\di^Dr\,\rho(\rv)$, and
\begin{equation}
v^{\rm LDA}_{\rm x}[\rho](\rv)\;\equiv\;\frac{\delta E^{\rm LDA}_{\rm x}[\rho]}{\delta\rho(\rv)}\;=\;
-A_D\Big(1+\frac1D\Big)\rho(\rv)^{1/D}.
\label{vxLDAasympt}
\end{equation}
In this case, since $v_{\rm SCE}[\rho](\rv)\to \frac{N-1}r$ and $v_{\rm H}[\rho](\rv)\to\frac{N}r$ for $r\equiv|\rv|\to\infty$,
we have asymptotically 
\begin{equation}
v_{\rm SCE}[\rho](\rv)-v_{\rm H}[\rho](\rv)\;\to\;-\frac1r\qquad(r\to\infty).
\label{asympt}
\end{equation}
Comparing this with Eq.~\eqref{vxLDAasympt}, we see that a solution $\rho(\rv)$ of Eq.~\eqref{EulerLam} must display
the asymptotic behavior
\begin{equation}
\rho(\rv)\to\frac{k_1}{r^D}\qquad(r\to\infty),
\label{eq_geddens}
\end{equation}
with some constant $k_1$. Such a function is evidently not normalizable, since 
with the $D$-dimensional volume element
$\di^Dr=k_2r^{D-1}\di r$ and a radius $R>0$ finite but large enough, we have
\begin{equation}
\int_{|\rv|\ge R}\di^Dr\,\rho(\rv)\;=\;\int_R^\infty\di r\frac{k_1k_2}r\;=\;\infty.
\end{equation}
We emphasize that this reasoning also applies to the modified functional
\begin{equation}
\widetilde{\Lambda}[\rho]\;=\;\frac{E_{\rm xc}[\rho]}{E^{\rm LDA}_{\rm x}[\rho]},
\end{equation}
where the indirect SCE interaction energy $V_{\rm ee}^{\rm SCE}[\rho]-U[\rho]$ is replaced with the functional
$E_{\rm xc}[\rho]$ of the exchange-correlation energy, since the xc potential for $N$-electron systems has the same
asymptotic behavior as Eq.~\eqref{asympt},
\begin{equation}
v_{\rm xc}[\rho](\rv)\;\equiv\;\frac{\delta E_{\rm xc}[\rho]}{\delta\rho(\rv)}\;\to\;-\frac1r\qquad(r\to\infty).
\end{equation}
Quite interestingly, a density of exactly the same form of Eq.~\eqref{eq_geddens} for the 3D case, but only restricted in a finite region of space (thus set to zero outside some region $r\in [R_1,R_2]$), has been considered by Perdew {\em et. al.} \cite{PerRuzSunBur-JCP-14} to study a general feature of GGA approximations related to the LO bound.
Notice, however, that if we consider this kind of densities,  $\rho(\rv)\propto r^{-3}$ in  $r\in [R_1,R_2]$, even by choosing $R_1$ and $R_2$ very large we get quite low values for $\Lambda[\rho]$, indicating that the asymptotic condition is anyway not enough to give a large $\Lambda$ value, see Table~\ref{tab:trialrho}.

Even more generally, in a fictitious universe where the electron-electron repulsion is multiplied by a factor
$\alpha\ge0$, the density functional of their xc energy is given by
\begin{equation}
E_{{\rm xc},\alpha}[\rho]=
\int_0^\alpha\di\beta\Big\{\big\langle\Psi_\beta[\rho]\big|\hat{V}_{\rm ee}\big|\Psi_\beta[\rho]\big\rangle-U[\rho]\Big\}.
\end{equation}
Here, out of all antisymmetric wave functions $\Psi$ that are associated with the same density $\rho$,
$\Psi_\beta[\rho]$ is the one that minimizes the expectaion $\langle\Psi|\hat{T}+\beta\hat{V}_{\rm ee}|\Psi\rangle$,
for any number $\beta\ge0$. Since the corresponding $\alpha$-dependent xc potential has the asymptotic behavior
\begin{equation}
v_{{\rm xc},\alpha}[\rho](\rv)\;\equiv\;\frac{\delta E_{{\rm xc},\alpha}[\rho]}{\delta\rho(\rv)}
\;\to\;-\frac{\alpha}r\qquad(r\to\infty),
\end{equation}
we conclude that even for the functional
\begin{equation}
\Lambda_\alpha[\rho]\;=\;\frac{\frac1\alpha E_{{\rm xc},\alpha}[\rho]}{E^{\rm LDA}_{\rm x}[\rho]},
\end{equation}
the maximizing function $\rho(\rv)$ must have compact support. Notice that $\Lambda_{\alpha=1}[\rho]=\widetilde{\Lambda}[\rho]$ and
$\lim_{\alpha\to\infty}\Lambda_\alpha[\rho]=\Lambda[\rho]$.

If $p(\rv)=0$ for $|\rv|\ge r_0$, we see that, in principle, a maximizing density in Eq.~\eqref{EulerLam_p} could exist. However, with our numerical investigation we have always found larger values of $\Lambda$ for densities with unbounded support.

\section{Following the functional gradient of $\Lambda[\rho]$}
\label{Following DG}
Although $\Lambda[\rho]$ has no maximizing density $\rho$ without compact support, the functional gradient $\delta\Lambda/\delta\rho$ tells
us how to increase the value $\Lambda[\rho]$ (or challenge the Lieb-Oxford bound) systematically.
Starting from  an $N$-electron density $\rho=\rho(\rv)$ with a high value
$\Lambda[\rho]$, we consider a small density variation,
\begin{equation}
\rho(\rv)\to\rho(\rv)+\eps\sigma(\rv),\qquad\int d^3r\,\sigma(\rv)=0.
\label{eq_sigma}
\end{equation}
Provided that $\eps\sigma(\rv)$ is truly ``small``, which precisely means that
\begin{equation}
\int d^3r\,\sigma(\rv)^2=1
\label{eq_sigmanorm}
\end{equation}
and $|\epsilon|\ll1$, we have
\begin{equation}
\Lambda[\rho+\eps\sigma]-\Lambda[\rho]\;\approx\;\eps\int d^3r\,G[\rho](\rv)\,\sigma(\rv),
\label{Gsig}
\end{equation}
with the gradient $G[\rho](\rv)\equiv\delta\Lambda[\rho]/\delta\rho(\rv)$ given by Eq.~\eqref{EulerLam}.
Although $\int d^3r\,\sigma(\rv)=0$, the right-hand side of Eq.~\eqref{Gsig} can nevertheless be $>0$, provided that
$G([\rho];\rv)$, as a function of $\rv$, is different from a constant, $G[\rho](\rv)\ne{\rm const}$.

\subsection{Formal optimization of the increment}
Formally, maximizing the integral $\int\di^3r\,G[\rho](\rv)\sigma(\rv)$ with respect to
$\sigma(\rv)$ subject to the two constraints $\int\di^3r\,\sigma(\rv)=0$ and $\int\di^3r\,\sigma(\rv)^2=1$,
\begin{eqnarray}
& & \frac{\delta}{\delta\sigma(\rv)}\Bigg\{\int\di^3r\,G[\rho](\rv)\sigma(\rv)
-\mu_1\int\di^3r\,\sigma(\rv) \nonumber \\
& & -\mu_2\int\di^3r\,\sigma(\rv)^2\Bigg\}=0,
\label{varLam}
\end{eqnarray}
yields the Euler Equation $G[\rho](\rv)-\mu_1-2\mu_2\sigma(\rv)=0$, with the solution
\begin{equation}
\sigma_0(\rv)=\frac{G[\rho](\rv)-\mu_1}{2\mu_2}.
\label{sig}
\end{equation}
The first Lagrange multiplier $\mu_1$ is fixed by the normalization constraint $\int d^3r\,\sigma_0(\rv)=0$.
The second one $\mu_2$ is absorbed in the small parameter $\eps$, guaranteeing the validity of the approximation
\eqref{Gsig}. [Independently, smallness of $\eps\sigma(\rv)$ is necessary (but not sufficient) for the
resulting density to be non-negative, $\rho(\rv)+\eps\sigma(\rv)\ge0$ for all $\rv$.]

For a $N$-electron (finite) density $\rho$, Eqs.~\eqref{EulerLam}-\eqref{asympt} imply the large-$r$ behavior ($r\to\infty$)
\begin{equation}
G[\rho](\rv)\;\to\;\frac1{|E^{\rm LDA}_{\rm x}[\rho]|}
\Big[\frac1r-\Lambda[\rho]A_D\Big(1+\frac1D\Big)\,\rho(\rv)^{1/D}\Big].
\label{eq_Gsigma0}
\end{equation}
Necessarily, $\sigma_0(\rv)\to 0$ for $r\to\infty$, implying $\mu_1=0$ in Eq.~\eqref{sig}. Consequently, due to the term $1/r$ in Eq.~\eqref{eq_Gsigma0}, $\int\di^3r\,\sigma_0(\rv)$ cannot be zero (or even finite). In other words, $\rho(\rv)+\eps\sigma_0(\rv)$, with $\eps\ne0$, must, again, yield a density with compact support. In the following, we give an analytical example for the case of a density with compact support.

\subsection{Analytical example for densities with compact support}
\label{app_analytic}
As an example, we evaluate Eq.~\eqref{sig} for the spherical 2-electron density \cite{RasSeiGor-PRB-11}
\begin{eqnarray}
\rho(\rv)=\left\{\begin{array}{cc}\rho_0 & (r\le R),\\ 0 & (r>R)\end{array}\right.\qquad\rho_0=\frac3{2\pi R^3}.
\end{eqnarray}
This density corresponds to a uniformly charged sphere (``droplet'') with radius $R$ and total charge 2,
\begin{eqnarray}
U[\rho]=\frac{12}{5R},\qquad
v_{\rm H}[\rho](r)=\frac{3R^2-r^2}{R^3} \qquad (r\le R).
\end{eqnarray}
The exact exchange energy $E_x[\rho]$ is given by $-E_x[\rho]=\frac12U[\rho]=\frac{1.2}R$, while \cite{RasSeiGor-PRB-11}
\begin{eqnarray}
& & -E_x^{\rm LDA}[\rho]=\frac{1.1545}{R},\\
& & v_x^{\rm LDA}[\rho](r)=-(\frac9{2\pi^2})^{1/3}\frac1R \qquad (r\le R).
\end{eqnarray}
From Ref.~\cite{RasSeiGor-PRB-11}, we have $\Lambda[\rho]=1.498$ and the SCE co-motion function (see Appendix~\ref{appSCE}) is
\begin{equation}
f(r)=R\Big(1-\frac{r^3}{R^3}\Big)^{1/3}.
\end{equation}
The resulting SCE external potential is given by
\begin{eqnarray}
& & v^{\rm SCE}[\rho](\rv)=v^{\rm SCE}[\rho](0)-\int_0^r\frac{du}{[u+f(u)]^2}\nonumber\\
& & = v^{\rm SCE}[\rho](0)-\frac1{R}\int_0^{r/R}\frac{dx}{[x+(1-x^3)^{1/3}]^2}.
\end{eqnarray}
Eventually, Eq.~\eqref{sig} reads
\begin{equation}
\sigma(r)=\frac{-v^{\rm SCE}[\rho](r)+v_{\rm H}[\rho](r)-\tilde{\mu}_1}{2\tilde{\mu}_2},
\end{equation}
where $\tilde{\mu}_1=\mu_1-\Lambda[\rho]v_x^{\rm LDA}$ (note that $v_x^{\rm LDA}$ does not depend on $\rv$ in
the present example) and $2\tilde{\mu}_2=-E_x^{\rm LDA}[\rho]2\mu_2>0$. The constant $v^{\rm SCE}[\rho](0)$ can also
be absorbed by the multiplier $\tilde{\mu}_1$ which is fixed by the conditon $\int_0^Rdr\,4\pi r^2\sigma(r)=0$.
Then, we have
\begin{equation}
\sigma(r)=\frac1{2\tilde{\mu}_2R}\Bigg\{3-\frac{r^2}{R^2}-\tilde{\mu}_1+\int_0^{r/R}\frac{dx}{[x+(1-x^3)^{1/3}]^2}\Bigg\}.
\end{equation}
A simple but accurate approximation to this function (for $R=1$) is
\begin{equation}
\sigma_{\rm appr}(r)=\frac1{2\tilde{\mu}_2}\Big[0.4r^3-1.85r^2+r+0.16\Big]\qquad r\le 1.
\end{equation}
We therefore consider the densities (for $r\le 1$)
\begin{equation}
\rho_a(r)=\rho_0+a\Big[0.4r^3-1.85r^2+r+0.16\Big]\qquad(a\ge0)
\end{equation}
to obtain the values $\Lambda[\rho_{0.2}]=1.521$, $\Lambda[\rho_{0.5}]=1.551$, $\Lambda[\rho_{1.0}]=1.590$, 
$\Lambda[\rho_{1.5}]=1.611$, $\Lambda[\rho_{1.6}]=1.612$. [For $a>1.6$, the density $\rho_a(r)$ becomes negative.]

\subsection{Compromise for $N$-representability for densities with unbounded support}
Since we observe higher values of $\Lambda[\rho]$ for densities for which $p(\rv)\neq 0$ everywhere, we consider here a compromise to follow the gradient of $\Lambda[\rho]$ without violating $N$-representability. We perturb the density with some function $\sigma(\rv)$ in Eq.~\eqref{eq_sigma} that depends on a certain number of parameters and satisfies Eqs.~\eqref{eq_sigma}-\eqref{eq_sigmanorm}, keeping the perturbed density $N$-representable with suitable constraints. One can then choose the parameter values in order to maximize the overlap with the gradient [the right-hand-side of Eq.~\eqref{Gsig}].

As an example, we start from the 3D exponential two-electron density ($D=3,\;N=2$)
\begin{equation}
\rho(\rv)\;=\;\frac{\ee^{-r}}{4\pi},   
\label{rhoExpl}
\end{equation}
which already gives the high value $\Lambda[\rho]=1.69905$ (see Table~\ref{tab:trialrho}).
We choose for $\sigma(\rv)$ the parametrized form
\begin{equation}
\sigma_a(\rv)\;=\;\sqrt{\frac{3a^3}\pi}\Big(1-\frac{ar}3\Big)\ee^{-ar}\qquad(a>0),
\label{sigExpl}
\end{equation}
which obeys the conditions $\int\di^3r\,\sigma_a(\rv)=0$ and $\int\di^3r\,\sigma_a(\rv)^2=1$ for all values of the
parameter $a>0$, so that the function $\rho_{a,\eps}(\rv)=\rho(\rv)+\eps\sigma_a(\rv)$ is always correctly normalized.
In addition, $N$-representability requires that $\rho_{a,\eps}(\rv)\ge0$ for all $r\ge0$.
For any value of $a$, this is fulfilled for $\eps_{\rm min}(a)\le\eps\le\eps_{\rm max}(a)$, where
$\eps_{\rm min}(a)\le0$ and $\eps_{\rm max}(a)\ge0$ are given by
\begin{eqnarray}
\eps_{\rm min}(a)=\left\{\begin{array}{c@{\quad\quad}c}
\frac{3(a-1)}{4a\sqrt{3a^3\pi}}\ee^{-(3-4a)/a}&(0<a\le\frac34),\\
-\frac1{4\sqrt{3a^3\pi}}&(a\ge\frac34),
\end{array}\right.
\end{eqnarray}
\begin{eqnarray}
\eps_{\rm max}(a)=\left\{\begin{array}{c@{\quad\quad}c}
0&(0<a\le1),\\
\frac{3(a-1)}{4a\sqrt{3a^3\pi}}\ee^{-(3-4a)/a}&(a\ge1).
\end{array}\right.
\end{eqnarray}

Evaluating numerically the functional gradient $G[\rho](\rv)\equiv\delta\Lambda[\rho]/\delta\rho(\rv)$ of
Eq.~\eqref{EulerLam} for the density $\rho(\rv)=\rho(r)$ of Eq.~\eqref{rhoExpl}, we consider, as a function of $a$,
the overlap integral
\begin{equation}
I(a)\;=\;\int_0^\infty\di r(4\pi r^2)\,G[\rho](r)\,\sigma_a(r).
\end{equation}
For any value of $a>0$, the maximum possible value in Eq.~\eqref{Gsig} is approximately (if the first-order expansion holds)
\begin{equation}
\Lambda\big[\rho+\eps(a)\sigma_a\big]-\Lambda[\rho]\;\approx\;\eps(a)\,I(a),
\label{LamOverlap}
\end{equation}
where $\eps(a)=\eps_{\rm max}(a)\ge0$ for $I(a)\ge0$ and $\eps(a)=\eps_{\rm min}(a)\le0$ for $I(a)\le0$.

Numerically, $I(a)>0$ for $0<a<1$ and $I(a)<0$ for $a>1$, with a strong maximum $I(a_1)\approx2.9267\cdot4\pi$ at
$a_1\approx0.079$ and a weak minimum $I(a_2)\approx-0.01479\cdot4\pi=-0.2302$ at $a_2\approx2.49$.
While $\eps(a_1)=0$, we have $\eps(a_2)=-0.0207$, and Eq.~\eqref{LamOverlap} for $a=a_2$ gives
\begin{equation}
\Lambda\big[\rho+\eps(a_2)\sigma_{a_2}\big]-\Lambda[\rho]\;\approx \;0.004~765.
\label{Lambda_a2}
\end{equation}
In Table~\ref{tab:gradvals} we report the values of $\Lambda[\rho+\epsilon\sigma_{a_2}]$ as a function of $\epsilon$ and compare them with the ones from the first-order expansion. As predicted, we see that $\Lambda[\rho]$ increases for small $\epsilon$. However, the first-order expansion breaks down before $\eps(a_2)$, so that the maximum value of $\Lambda[\rho]$ that we obtain is less than the one predicted by Eq.~\eqref{Lambda_a2}. The improvement in this case is very small, but we suspect that this is due to the fact that for $N=2$ the exact $\bar{\lambda}_3(2)$ is very close to 1.701, so that we are really hitting the boundary. In fact, in the previous example of Sec.~\ref{app_analytic}, we have seen that when we start from a much less optimal density the improvement in $\Lambda[\rho]$ with our procedure is much larger.

We have also repeated the procedure using as a starting density the one corresponding to $\epsilon=-0.02$ in Table \ref{tab:gradvals}, but we could only slightly improve the result obtaining $\Lambda[\rho]=1.701052$, which is, so far, our best value,
\begin{equation}
\bar{\lambda}_3(2)>1.701052.
\label{eq_lambda2final}
\end{equation}

\begin{table}
\begin{tabular}{|l||l|l|}\hline
$\;\;\;\eps$ & $\Lambda[\rho+\eps\sigma_{a_2}]$ & $\Lambda[\rho]+\eps I(a_2)$ \\ \hline\hline
$\;\;\;$0 & 1.699~052 & 1.699~052\\ \hline
$-$0.01 & 1.700~487 & 1.701~354\\ \hline
$-$0.015 & 1.700~833 & 1.702~505\\ \hline
$-$0.02 & 1.700~868 & 1.703~656\\ \hline
$-$0.0207 & 1.700~843 & 1.703~817\\ \hline
\end{tabular}

\caption{Exact values $\Lambda[\rho+\eps\sigma_{a_2}]$ for various values of $\eps<0$, compared with the first-order expansion.}
\label{tab:gradvals}
\end{table}

\section{Is a uniform density the most challenging for the Lieb-Oxford bound?}
\label{sec_unif}
In Ref.~\onlinecite{RasPitCapPro-PRL-09} it has been argued that the tightest bound should correspond to the case of the uniform electron gas at extremely low density (equivalent to the SCE limit for a uniform density). This suggestion was made by considering electronic hamiltonians of the form \eqref{eq:H} with particularly challenging $v(\rv)$, keeping in mind that the bound increases \cite{LieOxf-IJQC-81} with the number of electrons $N$, see Eq.~\eqref{eq_incN}. 

With our formalism, we directly consider the most challenging wavefunction (or one which is very close to it, thus providing anyway a lower bound for $\Lambda[\rho]$) for each given density, and we can thus question whether a uniform density profile is really the most challenging for the bound. Already by putting together existing data, we can compare, in Table~\ref{tab:atomsvsunif}, the values of $\Lambda[\rho]$ obtained from the (sphericalized) atomic densities of Li, Be, C, B, and Ne \cite{SeiGorSav-PRA-07}, with the ones obtained from spheres of uniform density (``droplets'') \cite{RasSeiGor-PRB-11}: we clearly see that the atomic densities yield significantly higher values of $\Lambda[\rho]$, as already observed for $N=2$ in Table~\ref{tab:trialrho}.
\begin{table}
\begin{tabular}{|c||c|c|}\hline
$N$ & $\Lambda[\rho]$ atomic & $\Lambda[\rho]$ droplet \\ \hline\hline
3 & 1.713 & 1.550\\ \hline
4 & 1.731 & 1.603\\ \hline
5 & 1.747 & 1.627\\ \hline
6 & 1.767 & 1.657\\ \hline
10 & 1.816 & 1.708\\ \hline
\end{tabular}
\caption{For different values of $N$, we compare the $\Lambda[\rho]$ obtained from atomic densities (values from \cite{SeiGorSav-PRA-07}) with the ones obtained from spherical droplets of uniform density (values from \cite{RasSeiGor-PRB-11}).}
\label{tab:atomsvsunif}
\end{table}
\begin{figure}
   \includegraphics[width=8cm]{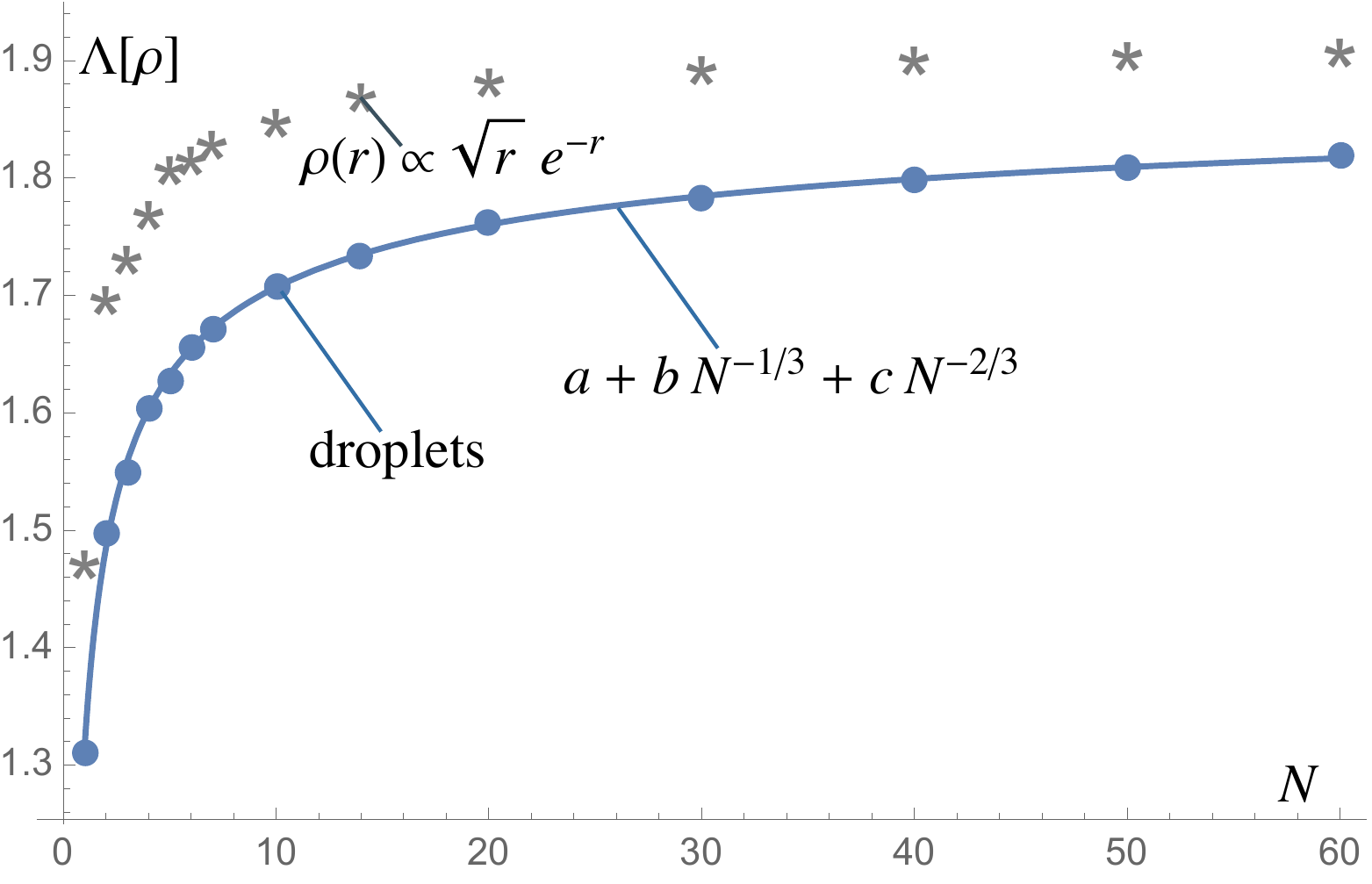}
   \caption{Values of $\Lambda[\rho]$ for the fixed density profile $\rho(r)\propto r^{1/2}e^{-r}$ compared with those for spheres of uniform density (``droplets'') as a function of the particle number $N$. The values for $\rho(r)\propto r^{1/2}e^{-r}$ are significantly higher than those for uniform densities. The size extrapolation for uniform densities is also shown, where the fitting parameters are $a=1.918$, $b=-0.3253$, $c=-0.2791$.}
\label{fig:LO}
\end{figure}

We have also performed calculations with the fixed spherical density profile $\rho(r)\propto r^{1/2}e^{-r}$, which was particularly promising for $N=2$ (see Table~\ref{tab:trialrho}), for particle numbers $N\le 60$, and compared the values with the ones for spheres of uniform density, extending the calculations of Ref.~\onlinecite{RasSeiGor-PRB-11} up to $N=60$. The results are reported in Fig.~\ref{fig:LO}, where we clearly see that the uniform droplets give values significantly lower for $\Lambda[\rho]$. This suggests that a similar behavior may arise in the limit $N\to\infty$: a density with particular modulations might challenge the bound more than the uniform one.

Our new value for the uniform sphere at $N=60$, $\Lambda=1.818$, sets an improved upper bound \cite{LewLie-PRA-15}, equal to $-1.343$,  for the low-density uniform electron gas indirect energy per particle $w$, which then must be between 
\begin{equation}
-1.45\le w\le -1.343,
\label{eq_jelllowdens}
\end{equation}
where the lower bound $-1.45$ has been proven in \cite{LieNar-JSP-75}. We have also performed a size extrapolation of our $\Lambda[\rho]$ for the droplets of uniform density of Fig.~\ref{fig:LO}, by fitting our data to a lquid-drop model expansion
\begin{equation}
\Lambda_{\rm unif}[N]=a+b \,N^{-1/3}+c\, N^{-2/3},
\end{equation}
finding $a=1.918$, $b=-0.3253$, $c=-0.2791$. The fitting function is also shown in Fig.~\ref{fig:LO}. The value of the fitted parameter $a$ gives our $N\to\infty$ extrapolation for $\Lambda[\rho]$ in the uniform electron gas, $\Lambda_{\rm unif}[N\to\infty]=1.918$. This value can be compared with the one obtained by taking the $r_s\to\infty$ limit of popular LDA parametrizations: for example, the PW92 \cite{PerWan-PRB-92} parametrization yields  1.947 at zero polarization and 1.977 for the fully polarized case, while the VWN \cite{VosWilNus-CJP-80} at zero polarizations gives 1.9043.

After Lewin and Lieb \cite{LewLie-PRA-15} showed that the value $1.955=1.4442/A_3$ does not correspond to an indirect energy, our value $\Lambda[\rho]=1.91175$ for $N=60$ and spherically-symmetric density profile $\rho(r)\propto r^{1/2}e^{-r}$ is the highest value of $\lambda_3[\Psi]$ ever observed, setting a new lower bound for $\bar{\lambda}_3(N)$ for any $N$, so that, rigorously
\begin{equation}
\label{eq_newlower}
1.91175 \le \bar{\lambda}_3 \le 2.215,
\end{equation}
or, in terms of the constant $C_3$ in Eq.~\eqref{LOorg}
\begin{equation}
1.4119 \le C_3 \le 1.6358.
\label{eq_newC}
\end{equation}

\section{Conclusions and Perspectives}
\label{sec_conc}
We have developed a method to maximally challenge the Lieb-Oxford bound, using optimal (or nearly optimal) trial wave functions that can be constructed from a given density. This allows us to rewrite the most challenging bound for a given number of particles directly as a density functional. As a first application of the method, 
\begin{itemize}
	\item we improved -- see Eq.~\eqref{eq_lambda2final} -- the constant in the LO bound for $N=2$, which provides a constraint to develop new metaGGA functionals \cite{SunRuzPer-PRL-15}; 
	\item we have given an improved lower bound for the constant appearing in the LO inequality valid for all particle numbers $N$, see Eqs.~\eqref{eq_newlower}-\eqref{eq_newC};
	\item we have obtained an improved upper bound for the indirect energy per particle of the low-density uniform electron gas, see Eq.~\eqref{eq_jelllowdens}.
\end{itemize}
In future works we will analyze systematically the bound for larger particle numbers $N$, trying to give improved lower bounds for $\bar{\lambda}_D(N)$ and for $\bar{\lambda}_D$. 

More generally, from this study we have learned that it is quite difficult to predict which densities will maximally challenge the bound (see for example Table~\ref{tab:trialrho}: the trends reported there seem totally unpredictable). For sure, we observe that, for finite $N$, a uniform density is not the one that challenges the bound the most, suggesting that the indirect energy of the uniform gas at low-density may not provide the tightest bound, contrary to what was previously suggested.

\acknowledgments
We are really happy to dedicate this paper to Andreas Savin. His curiosity, deep knowledge, integrity, creativity and generosity are an infinite source of inspiration. 

We are very grateful to Lukas Schimmer, Mathieu Lewin, Kieron Burke and John Perdew for a critical reading of the manuscrpt, providing several suggestions to improve the paper. Financial support from the European Research Council under H2020/ERC Consolidator Grant ``corr-DFT'' (Grant No. 648932) and the Netherlands Organization for Scientific Research (NWO) through an ECHO grant (717.013.004) is acknowledged.

\appendix
\section{Strictly correlated electrons}
\label{appSCE}

The minimizing antisymmetric wave function $\Psi=\Phi_{\rm KS}[\rho]$ in the definition of the density
functional of the non-interacting kinetic energy \cite{Lev-PNAS-79},
\begin{equation}
T_{\rm s}[\rho]\;=\;\min_{\Psi\to\rho}\langle\Psi|\hat{T}|\Psi\rangle,
\end{equation}
is usually a Slater determinant of Kohn-Sham orbitals. In contrast, the minimizing one in
\begin{equation}
V_{\rm ee}^{\rm SCE}[\rho]\;=\;\min_{\Psi\to\rho}\langle\Psi|\hat{V}_{\rm ee}|\Psi\rangle
\label{eq_SCEprob}
\end{equation}
is (or it is very close to) a state $\Psi=\Psi_{\rm SCE}[\rho]$ with strictly correlated eletrons (SCE). $\Psi_{\rm SCE}[\rho]$ is not a
regular wave function but a Dirac-type distribution. Its position representation is singular,
\begin{equation}
\Psi^{\rm SCE}[\rho](\rv_1\sigma_1,...,\rv_N\sigma_N)=0\quad\mbox{for}\quad(\rv_1,...,\rv_N)\notin\Omega_0[\rho].
\end{equation}
Here, $\Omega_0[\rho]$ is a $D$-dimensional subspace of the $(N\times D)$-dimensional configuration space of the
$N$-electron system, given by
\begin{equation}
\Omega_0[\rho]\;=\;\Big\{\big(\sv,\fv_2(\sv),...,\fv_N(\sv)\big)\;\Big|\;\sv\in S_\rho\subset\R^D\Big\},
\end{equation}
where $S_\rho=\{\rv\in\R^D\,|\,\rho(\rv)\ne0\}$ is the spatial region of non-zero density. $\fv_1(\sv)\equiv\sv$,
$\fv_2(\sv)$, ..., $\fv_N(\sv)$ are co-motion functions: In an SCE state, a configuration $(\rv_1,...,\rv_N)$ is
observable only when its $N$ positions obey the relations, $\rv_n=\fv_n(\rv_1)$, $n=2,...,N$. 
Then, the distance between electrons $i$ and $j$ is $|\fv_i(\sv)-\fv_j(\sv)|$, fixed by the position $\rv_1=\sv$
of electron 1. Consequently, we have
\begin{equation}
V_{\rm ee}^{\rm SCE}[\rho]\;=\;
\int\di^Ds\,\frac{\rho(\sv)}N\sum_{i=1}^{N-1}\sum_{j=i+1}^N\frac1{|\fv_i(\sv)-\fv_j(\sv)|}.
\label{VeeSCEfs}
\end{equation}
This is truly a density functional, since the co-motion functions are fixed by the density,
$\fv_n(\sv)=\fv_n[\rho](\sv)$.
For a large class of densities $\rho$, the functions $\fv_n[\rho](\sv)$ and thus the functional
$V_{\rm ee}^{\rm SCE}[\rho]$ can be evaluated rigorously \cite{Sei-PRA-99,SeiGorSav-PRA-07,ButDepGor-PRA-12,ColDepDiM-CJM-14}. Its functional derivative turns out to be \cite{MalGor-PRL-12,MalMirCreReiGor-PRB-13}
\begin{equation}
\frac{\delta V_{\rm ee}^{\rm SCE}[\rho]}{\delta\rho(\rv)}\;=\;v_{\rm SCE}[\rho](\rv),
\end{equation}
with the SCE external potential $v_{\rm SCE}[\rho](\rv)$, fixed by
\begin{equation}
\nabla v_{\rm SCE}[\rho](\rv)\;=\;-\sum_{n=2}^N\frac{\rv-\fv_n(\rv)}{\,|\rv-\fv_n(\rv)|^3}.
\label{NablaVsce}
\end{equation}
As usual, the functional derivative is determined up to a constant, which for finite systems we fix by requiring that the potential vanishes at infinity.

For example, a spherical two-electron density $\rho(\rv)=\rho(r)$ in $D$ dimensional space
has the co-motion function \cite{Sei-PRA-99,ButDepGor-PRA-12}
\begin{equation}
\fv_2(\sv)\;=\;-f(s)\frac{\sv}{|\sv|}.
\label{ftwo}
\end{equation}
In terms of the invertible function
\begin{equation}
N_e(s)\equiv\int_{|\rv|\le s}\di^Dr\,\rho(r),
\label{Netwo}
\end{equation}
the radial co-motion function is given by
\begin{equation}
f(s)\;=\;N_e^{-1}\big(2-N_e(s)\big).
\end{equation}
Eq.~\eqref{ftwo} implies that $|\rv-\fv_2(\rv)|=r+f(r)$ and, due to Eq.~\eqref{VeeSCEfs},
\begin{equation}
V_{\rm ee}^{\rm SCE}[\rho]\;=\;\frac12\int\di^Dr\,\frac{\rho(r)}{r+f(r)}.
\label{VeeSCEtwo}
\end{equation}
Due to Eq.~\eqref{NablaVsce}, the SCE external potential, with $v_{\rm SCE}\to0$ for $r\to\infty$, is
\begin{equation}
v_{\rm SCE}[\rho](r)\;=\;\int_r^\infty\frac{\di s}{[s+f(s)]^2}.
\end{equation}

For any $N$-electron density $\rho(\rv)$ with co-motion functions $\fv_n(\rv)$ $(n=2,...,N)$, we may consider the
continuous series of scaled $N$-electron densities $\rho_\xi(\rv)=\xi^D\rho(\xi\rv)$, with $\xi>0$ and 
$\int\di^Dr\,\rho_\xi(\rv)=\int\di^Dr\,\rho(\rv)=N$. The co-motion functions $\fv^{(\xi)}_n(\rv)$ of $\rho_\xi(\rv)$
are given by
\begin{equation}
\fv^{(\xi)}_n(\rv)\;=\;\frac1\xi\fv_n(\xi\rv).
\end{equation}
Therefore, the functional \eqref{VeeSCEfs} has the simple scaling property
\begin{equation}
V_{\rm ee}^{\rm SCE}[\rho_\xi]\;=\;\xi V_{\rm ee}^{\rm SCE}[\rho].
\label{VeeSCEscal}
\end{equation}

We should remark that the SCE wave function as a minimizer for the electron-electron interaction energy has been first conjectured on physical grounds \cite{Sei-PRA-99,SeiGorSav-PRA-07}. In recent years, it was recognized that the problem posed by the minimization \eqref{eq_SCEprob} is equivalent to an optimal transport problem with Coulomb cost \cite{ButDepGor-PRA-12,CotFriKlu-CPAM-13}.
Since then, the optimal transport community has produced several rigorous results. In particular, the SCE state has been proven to be the true minimizer for any $N$ in 1D \cite{ColDepDiM-CJM-14} and in any dimension for $N=2$ \cite{ButDepGor-PRA-12}. For more general cases, it has been shown that the minimizer might not be of the SCE form \cite{ColStr-arxiv-15}. Even in that case, however, SCE-like solutions seem to be able to go very close to the true minimum \cite{DiMNenGor-XXX-15}.


\end{document}